\documentclass[twocolumn,superscriptaddress,prl,longbibliography,nobalancelastpage,aps]{revtex4-2}
\usepackage[pdftex, pdftitle={Reply to Pelliciari}, pdfauthor={Charles Tam}, colorlinks,allcolors=blue]{hyperref}
\usepackage{graphicx}
\usepackage{booktabs}
\usepackage{amsmath}
\usepackage{multirow}
\usepackage{array}

\begin{document}

\title{Reply to ``Comment on newly found Charge Density Waves in infinite layer Nickelates''}

\author{Charles C. Tam}
    \thanks{These authors contributed equally}
    \affiliation{Diamond Light Source, Harwell Campus, Didcot OX11 0DE, United Kingdom.}
    \affiliation{H. H. Wills Physics Laboratory, University of Bristol, Bristol BS8 1TL, United Kingdom.}

\author{Jaewon Choi}
    \thanks{These authors contributed equally}
    \affiliation{Diamond Light Source, Harwell Campus, Didcot OX11 0DE, United Kingdom.}
    
\author{Xiang Ding}
    \thanks{These authors contributed equally}
    \affiliation{School of Physics, University of Electronic Science and Technology of China, Chengdu, 610054, China}

\author{Stefano Agrestini}
    \affiliation{Diamond Light Source, Harwell Campus, Didcot OX11 0DE, United Kingdom.}

\author{Abhishek Nag}
    \affiliation{Diamond Light Source, Harwell Campus, Didcot OX11 0DE, United Kingdom.}
    \affiliation{Laboratory for Non-linear Optics, Paul Scherrer Institut, CH-5232 Villigen, Switzerland}
    
\author{Mei Wu}
    \affiliation{International Center for Quantum Materials and Electron Microscopy Laboratory, School of Physics, Peking University, Beijing 100871, China}

\author{Bing Huang}
    \affiliation{Beijing Computational Science Research Center, Beijing 100193, China}

\author{Huiqian Luo}
    \affiliation{Beijing National Laboratory for Condensed Matter Physics, Institute of Physics, Chinese Academy of Sciences, Beijing 100190}
    \affiliation{Songshan Lake Materials Laboratory, Dongguan, Guangdong 523808, China}
    
\author{Peng Gao}
    \affiliation{International Center for Quantum Materials and Electron Microscopy Laboratory, School of Physics, Peking University, Beijing 100871, China}

\author{Mirian Garc\'ia-Fern\'andez}
    \affiliation{Diamond Light Source, Harwell Campus, Didcot OX11 0DE, United Kingdom.}

\author{Liang Qiao}
    \email{liang.qiao@uestc.edu.cn}
    \affiliation{School of Physics, University of Electronic Science and Technology of China, Chengdu, 610054, China}

\author{Ke-Jin Zhou}
    \email{kejin.zhou@diamond.ac.uk}
    \affiliation{Diamond Light Source, Harwell Campus, Didcot OX11 0DE, United Kingdom.}

% \date{\today}

\begin{abstract}
Charge density waves (CDW) have been reported in NdNiO$_2$ and LaNiO$_2$ thin films grown on SrTiO$_3$ substrates using Ni-$L_3$ resonant x-ray scattering in Refs.~\cite{Tam_NM_2022,Rossi_NP_2022,Krieger_PRL_2022}. In their comment~\cite{Pelliciari__2023} on these reports, Pelliciari \textit{et al.} found no evidence for a CDW in a NdNiO$_2$ film by performing fixed-momentum energy-dependent measurements. Instead, they observed a nearby non-resonant scattering peak, attributed to the (101) substrate reflection, made accessible at Ni-$L_3$ due to third harmonic light contamination. Here we present fixed-momentum energy-dependent resonant inelastic x-ray scattering measurements across Ni-$L_3$ on NdNiO$_2$, used in the preceding study~\cite{Tam_NM_2022}. We see intrinsic Ni-$L_3$ energy profiles at all measured \textbf{Q} values, including a strong resonance effect at $\mathbf{Q}_\mathrm{CDW} = (-1/3, 0, 0.316)$ reciprocal lattice units. Attempts to measure the (101) substrate peak using third harmonic light at Ni-$L_3$ at I21, Diamond were unfruitful. Our results clearly demonstrate the electronic origin of the scattering peak published in Ref.~\cite{Tam_NM_2022} and lack of a detectable structural component in the peak.
\end{abstract}

\maketitle

\section{Introduction}

The infinite-layer nickelate superconductors contain-two-dimensional NiO$_2$ layers, nominal $S=1/2$ $3d^9$ Ni$^{1+}$ ions and an active $d_{x^2-y^2}$ orbital near the Fermi level, making them in many ways analogous to cuprate superconductors. Despite the absence of long-range antiferromagnetic ordering, highly dispersive magnons confirm the existence of strong electronic interactions~\cite{Lu_S_2021}. Strong electronic correlations often give rise to symmetry-breaking orders, which was confirmed in measurements of CDWs in LaNiO$_2$~\cite{Rossi_NP_2022}, NdNiO$_2$~\cite{Krieger_PRL_2022,Tam_NM_2022}, and PrNiO$_2$~\cite{Ren__2023}.

In Refs.~\cite{Tam_NM_2022,Krieger_PRL_2022,Rossi_NP_2022,Ren__2023}, scattering peaks were found around $\mathbf{Q}_\mathrm{CDW} = (1/3, 0,L)$ reciprocal lattice units (where $L\sim 0.3$) with a clear resonance effect around the Ni-$L_3$. In our previous work on NdNiO$_2$~\cite{Tam_NM_2022}, the resonant effect (Fig.~1e, Ref.~\cite{Tam_NM_2022}) and polarisation dependence (Figs.~S11, S12, Ref.~\cite{Tam_NM_2022}) suggest the observed scattering peak originates from charge correlations rather than spin correlations or structure. Energy dependence around the Ni-$L_3$ absorption edge were performed at a fixed scattering angle $2\theta$ (defined as $\Omega$ hereafter).

With a fixed $\Omega$, the total momentum transfer $\mathbf{Q}$ varies with $E$ as $\mathbf{Q} = (4\pi/hc)E \sin(\Omega/2)$ where $h$ is Planck's constant and $c$ is the speed of light. To fix the in-plane momentum transfer (\textit{i.e.} $Q_{H,K} = 1/3$), the incident angle $\theta$ needs to be changed as $Q_{H,K} = Q \sin(\theta-\Omega/2)$. This leaves the out-of-plane momentum transfer $Q_L = Q\cos(\theta-\Omega/2)$ variable. We will explain at the end of the reply that this type of measurement is valid if the CDW is sufficiently broad in $L$. 

In their comment, Pelliciari \textit{et al.} performed resonant elastic x-ray scattering (REXS) measurements on a NdNiO$_2$ film grown on a SrTiO$_3$ (STO) substrate~\cite{Pelliciari__2023}. For the sake of the reader we will briefly summarise their comment, referring to their main results as \textbf{(A)} and \textbf{(B)}.

\subsection{(A) Lack of resonance effect near $\mathbf{Q}_\mathrm{CDW}$ in NdNiO$_2$}

When doing energy dependence with a fixed scattering angle (now referred to as $E_{\mathrm{fix}\Omega}$), Pelliciari \textit{et al.} observed an enhancement centred around the Ni-$L_3$ edge. However, an energy dependent measurement at a fixed momentum transfer (now referred to as $E_{\mathrm{fix}Q}$) at $\mathbf{Q} = (1/3, 0, 0.31)$ yielded no resonance effect. They explored reciprocal space and found the (101) reflection of the STO substrate in proximity, made accessible at Ni-$L_3$ by third harmonic contamination \textit{i.e.}, making (101) appear at $(1/3, 0, 1/3)$.

\subsection{(B) Observation of $(1/3, 0, 1/3)$ on SrTiO$_3$ at Ni-$L_3$}

To confirm this, Pelliciari \textit{et al.} directly measured the STO substrate material. Similar to result \textbf{(A)}, the $E_{\mathrm{fix}\Omega}$ scan shows a resonance effect, while the $E_{\mathrm{fix}Q}$ scan does not. Based on this, Pelliciari \textit{et al.} concluded that scattering at $\mathbf{Q} = (1/3, 0, 0.31)$ contains substantial contamination from the substrate from higher order light. They proposed a \textit{new standard} procedure $E_{\mathrm{fix}Q}$ scans at different \textbf{Q} locations to determine the resonant contribution to the quasi-elastic scattering.

In this reply, we conduct resonant inelastic x-ray scattering (RIXS) experiments, including the proposed \textit{new standard} measurement, to address results \textbf{(A)} and \textbf{(B)}. We show in the experimental conditions of Ref.~\cite{Tam_NM_2022}, charge scattering is indeed observed at the CDW position, with negligible scattering originating from the substrate.

\section{Results and discussion}

\subsection{Reply to (A) - Observation of resonance effect at $\mathbf{Q}_\mathrm{CDW}$ on NdNiO$_2$ film}

First, we self consistently determined the scattering peak position in 3D reciprocal space. To do this, we placed the incident energy at the main peak of Ni$^{1+}$ in the x-ray absorption spectra (XAS) at $E_i=853\;$eV, where the scattering peak was seen to resonate~\cite{Tam_NM_2022,Krieger_PRL_2022}. Plotted in Fig.~\ref{fig1}a is an $H$ scan at fixed $L$, and in Fig.~\ref{fig1}b an $L$ scan at fixed $H$. We find the scattering peaked at $\mathbf{Q}_\mathrm{CDW} = (-0.337, 0, 0.316)$. 

With the exact value of $\mathbf{Q}_\mathrm{CDW}$ determined, we then moved onto the \textit{new standard} $E_{\mathrm{fix}Q}$ measurements. Energy scans across the Ni-$L_3$ edge were performed at three fixed \textbf{Q} positions. Off the peak and at a strictly non-resonant location $\mathbf{Q} = (-0.2, 0, 0.316)$ (Fig.~\ref{fig1}c), the measurement is purely fluorescence and resembles closely XAS. This indicates the presence of Ni, which is an intrinsic signal from the film. On the peak at $\mathbf{Q} = (-0.337, 0, 0.316)$ (Fig.~\ref{fig1}e), the signal becomes clearly resonant, as the energy profile is distinct from XAS and the overall intensity is $\sim \times 100$ stronger. Slightly away from the peak, at $\mathbf{Q} = (-0.29, 0, 0.316)$, (Fig.~\ref{fig1}d), the energy profile is still distinct from XAS, albeit with less intensity than on the peak, demonstrating the finite correlation length in $H$ ($\xi_H\approx 60\;\mathrm{\AA}$~\cite{Tam_NM_2022}).

The $E_{\mathrm{fix}Q}$ scans clearly show an evolution from the fluorescence-like signal off the peak to the strong resonant behaviour on the peak, in contrast to the non-resonant profile on the peak in Ref.~\cite{Pelliciari__2023}. From this, we conclude the $\mathbf{Q} = (-0.337, 0, 0.316)$ scattering peak has electronic origin.

\subsection{Reply to (B) - No evidence of third harmonic Ni-$L_3$ contamination at I21, Diamond}

Next we address the issue of third harmonic contamination at beamline I21, Diamond, by searching for the (101) reflection directly on the STO substrate material using third harmonic light. To begin, we started with the conditions where the CDW has been most frequently reported, namely with $E_i = 853\;$eV, at the resonance of the Ni$^{1+}$ XAS peak, at fixed $\Omega=154^\circ$, and scanned along $(H,0)$. Only a background-like signal was seen (Fig.~\ref{fig1}f). To investigate further, we went to $E_i=767\;$eV, where the third harmonic light (2301$\;$eV) fulfils exactly the (101) diffraction condition of STO when $\Omega = 154^\circ$. Again, only a background-like signal was seen (Fig.~\ref{fig1}g). 

Independent of this measurement, we examined the theoretical transmission of each optical component at the I21 beamline, and obtained a total transmission ratio of third harmonic to the primary beam at 853\;eV to be around $2.3\times10^{-10}$ ~\cite{Zhou_JoSR_2022}. Details of the calculation are in the Methods section. It is clear that the third harmonic contamination at the Ni $L_3$ edge at I21, Diamond is negligible, resulting in the lack of detection of the (101) substrate peak.

\subsection{E$_\mathrm{fixQ}$ vs E$_{\mathrm{fix}\Omega}$}

\begin{figure*}[!htb]
\center{\includegraphics{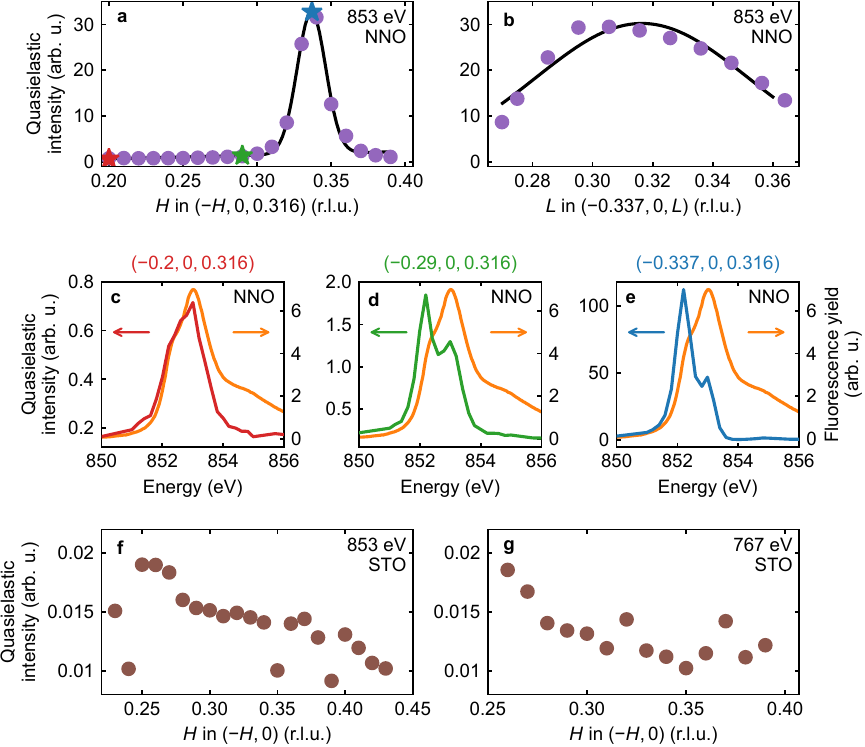}}
\caption{\textbf{a} Quasi-elastic RIXS intensity of NdNiO$_2$, along $H$ at $L=0.316$, taken with incident energy at the maximum of the Ni-$L_3$ absorption peak at 853\;eV. \textbf{b} Quasi-elastic RIXS intensity of NdNiO$_2$, along a small $L$ range limited by soft x-ray RIXS, at $H=-0.337$, taken at 853\;eV. Black lines are Gaussian fits. \textbf{c}-\textbf{e} Fixed \textbf{Q} energy scans of NdNiO$_2$, taken at fixed $\mathbf{Q} = (-0.2, 0, 0.316)$, $(-0.29, 0, 0.316)$ and $(-0.337, 0, 0.316)$, respectively. Fluorescence yield XAS taken with $\sigma$ polarisation at grazing incidence are plotted on the secondary axis. \textbf{f} $H$ scan of SrTiO$_3$ taken at 853\;eV. \textbf{g} $H$ scan of SrTiO$_3$ taken at 767\;eV. All measurements were conducted at 20\;K with $\sigma$ polarised x-rays.}
\label{fig1}
\end{figure*}

Finally, we comment on the validity of the $E_{\mathrm{fix}\Omega}$ scan. For the $E_{\mathrm{fix}\Omega}$ scan employed in the previous study at $\Omega = 154^\circ$, the out-of-plane momentum transfer $Q_L$ was changed from 0.365 to 0.368\;r.l.u. in the energy window 851 to 855\;eV~\cite{Tam_NM_2022}. Comparing to the FWHM of the CDW (0.08\;r.l.u.) in $L$ (Fig~\ref{fig1}b), $\Delta Q_{L}$ (0.003\;r.l.u.) is significantly smaller. In this case, although the CDW has clear $L$ dependence, it is broad enough so that $E_{\mathrm{fix}\Omega}$ is a sufficient approximation of $E_{\mathrm{fix}Q}$.  

\section{Conclusion}
We reexamined the CDW in the same sample, using the same RIXS instrument as in Ref.~\cite{Tam_NM_2022}. Following the determination of the exact $\mathbf{Q}_\mathrm{CDW}$, we performed the \textit{new standard} procedure suggested by Pelliciari \textit{et al.}, namely $E_{\mathrm{fix}Q}$ scans at various \textbf{Q} locations. This revealed a distinct evolution of energy profiles from off to on the CDW peak, in contrast to results of Pelliciari \textit{et al.} Searching for the (101) reflection on a bare piece of the STO substrate material using third harmonic light at Ni-$L_3$ yielded no results. 

These results demonstrate that the scattering peak observed in our nickelate films are primarily electronic in origin, with a negligible structural part. Results shown by Pelliciari \textit{et al.} are self-consistent but only demonstrate the lack of charge correlations in their nickelate sample, while higher amounts of third harmonic contamination at the REXS instrument used allow for the detection of the (101) substrate reflection. Therefore, their interpretation cannot be validly applied to data collected in Refs.~\cite{Tam_NM_2022,Rossi_NP_2022,Krieger_PRL_2022}.

\section{Methods}
A nickelate film sample (called NNO-1 in Ref.~\cite{Tam_NM_2022}) used in our previous study was chosen for the current measurement~\cite{Tam_NM_2022}. Samples were grown on SrTiO$_3$ substrates by pulsed laser deposition followed by topotactic reduction with CaH$_2$. The sample does not have a capping layer. More sample growth and characterisation details can be found in Ref.~\cite{Tam_NM_2022}.

We performed RIXS measurements at the I21 beamline, Diamond Light Source, UK, where previously published nickelate CDW data were collected~\cite{Tam_NM_2022,Rossi_NP_2022}. All measurements were performed at 20\;K with an energy resolution of 37.2\;meV (FWHM). The sample was aligned with the same geometry as previously, namely the crystallographic a–c plane aligned with the horizontal scattering plane. To maximise the CDW signal we used $\sigma$ polarisation and a grazing-in geometry ($\theta < \Omega/2$ as shown in Ref~\cite{Tam_NM_2022} Fig.~S8). $\Omega$ was fixed at $154^\circ$ in the $E_{\mathrm{fix}\Omega}$ geometry. Reciprocal space is labelled reciprocal lattice units (r.l.u.) of the tetragonal structure of NdNiO$_2$, using refined lattice parameters $a=b=3.908\;\mathrm{\AA}$ and $c=3.543\;\mathrm{\AA}$. 

For the estimation of the third harmonic transmission of the beamline, the efficiency of each optical element was calculated by either reflectivity (for mirrors) or a diffraction calculation (for gratings). Details of the optical elements can be found in Ref.~\cite{Zhou_JoSR_2022}. The results of these calculations are listed in Table.~\ref{tab1}. The transmission ratio of the beamline (up to and including M4) is estimated to be $2.4\times10^{-6}$. The spectrometer (M5 and SVLS2) cuts down the transmission by a further $9.8\times10^{-5}$, resulting in a total transmission of third harmonic light at Ni-$L_3$ of around $2.3\times10^{-10}.$

\begin{table*}[!htb]
\begin{tabular}{ccccc}
\cmidrule{1-4}

\textbf{Optical element}         &  \textbf{Primary\;(853\;eV)}  &  \textbf{3rd harmonic\;(2559\;eV)}  &  \textbf{3rd/1st ratio} \\ 

\cmidrule{1-4}
ID flux (photons$\;$sec$^{-1}$) & $2.24\times10^{15}$       & $6.1\times10^{14}$               & 0.27  &               \multirow{6}{*}{\hspace{-1em}$\left.\begin{array}{l}
                                                                                                                        \\
                                                                                                                        \\
                                                                                                                        \\
                                                                                                                        \\
                                                                                                                        \\
                                                                                                                        \\
                                                                                                                        \end{array}\right\rbrace 2.4\times10^{-6}$} \\
M1 reflectivity (C coating)      & 0.92                      & 0.93                             & 1       &             \\
M2 reflectivity (C coating)      & 0.89                      & 0.1                              & 0.11    &             \\
M3 reflectivity (Pt coating)     & 0.59                      & 0.35                             & 0.59    &             \\
VPG2 grating reflectivity        & 0.12                      & $2.7\times10^{-5}$               & $2.25\times10^{-4}$ & \\
M4 reflectivity (Pt coating)     & 0.8                       & 0.49                             & 0.6          &        \\
M5 figure of merit (Pt coating)  & 0.028                     & 0.0028                           & 0.1         &         \multirow{2}{*}{\hspace{-1em}$\left.\begin{array}{l}
                                                                                                                        \\
                                                                                                                        \\
                                                                                                                        \end{array}\right\rbrace 9.8\times10^{-5}$} \\
SVLS2 grating reflectivity       & 0.045                     & $4.45\times10^{-5}$              & $9.8\times10^{-4}$    \\ 
\cmidrule{1-4}
\end{tabular}
\caption{Photon flux and calculated efficiency of optical components for Ni-$L_3$ (853\;eV) and third harmonic light (2559\;eV) of optical components in use at I21, Diamond (at the time of writing). The beamline transmission ratio is $2.4\times10^{-6}$, while the spectrometer is $9.8\times10^{-5}$ giving a total transmission ratio of third harmonic to primary light of $2.3\times10^{-10}$.}
\label{tab1}
\end{table*}

\section{Acknowledgements}

We acknowledge fruitful discussions with Giacomo Ghiringhelli, Wei-Sheng Lee and Mark Dean. All data were taken at the I21 RIXS beamline of Diamond Light Source (United Kingdom) using the RIXS spectrometer designed, built and owned by Diamond Light Source.  We acknowledge T. Rice for the technical support throughout the experiments. C.C.T. acknowledges funding from Diamond Light Source and the University of Bristol under joint doctoral studentship STU0372. L.Q. and H.L. thank the support from the NSFC (Grant Nos. 11774044, 52072059 and 11822411) and SPRP-B of CAS (Grant No. XDB25000000). K.-J.Z. and H.L. thank the support from NSF of Beijing (Grant No. JQ19002).

\section{Competing interests}

The authors declare no competing interests.

\section{Data availability}

Relevant data are available from the corresponding authors upon reasonable request.

\bibliography{refs}
\end{document}